\documentclass[11pt]{article}
\usepackage{latexsym}
\usepackage{graphicx}
\usepackage{mathptmx}
\usepackage{multirow}

\usepackage{amsthm}
\usepackage{comment}
\usepackage{amsmath}
\usepackage{amssymb}
\usepackage{amsbsy}
\usepackage{amsfonts}
\DeclareMathAlphabet{\mathcal}{OMS}{cmsy}{m}{n}

\usepackage{geometry}
\usepackage[pdftex,colorlinks=true,urlcolor=blue,citecolor=black,anchorcolor=black,linkcolor=black]{hyperref}
\usepackage{algorithm}
\usepackage{algpseudocode}
\usepackage{algorithmicx}

\algdef{SE}[FOR]{ForEach}{EndForEach}[1]{\algorithmicfor\ #1\ \algorithmicdo}{\algorithmicend\ \algorithmicfor}
\begin{document}

\title{SF-SFD: Stochastic Optimization of Fourier Coefficients to Generate Space-Filling Designs}

\author{Manisha Garg\\[12pt]
    Department of Mathematics\\
    University of Illinois Urbana-Champaign\\
    Urbana, IL, USA 61801\\
\and
Tyler H. Chang\\[12 pt]
    Mathematics and Computer Science Division\\
	Argonne National Laboratory\\
    Lemont, IL, USA 60439 \\
\and
Krishnan Raghavan\\[12pt]
    Mathematics and Computer Science Division\\
	Argonne National Laboratory\\
    Lemont, IL, USA 60439 \\
}

\date{Dec 2023}

\maketitle

\section*{Abstract}
Due to the curse of dimensionality, it is often prohibitively expensive to generate deterministic space-filling designs. On the other hand, when using na{\"i}ve uniform random sampling to generate designs cheaply, design points tend to concentrate in a small region of the design space. Although, it is preferable in many cases to utilize quasi-random techniques such as Sobol sequences and Latin hypercube designs over uniform random sampling, these methods have their own caveats especially in high-dimensional spaces. In this paper, we propose a technique that addresses the fundamental issue of measure concentration by updating high-dimensional distribution functions to produce better space-filling designs. Then, we show that our technique can outperform Latin hypercube sampling and Sobol sequences by the discrepancy metric while generating moderately-sized space-filling samples for high-dimensional problems.

\section{Introduction}
\label{sec:intro}

Design of experiments is a critical first step in numerous application areas including
statistical response surface methodology (RSM), surrogate-based optimization, and
modeling of complex systems \cite{myers2016}.
To give a few examples, deterministic, random, and quasi-random experimental designs are generally applied in engineering applications ranging from particle accelerator designs~\cite{neveu2022}, to high-performance computing~(HPC) performance analysis \cite{wang2023}, and
generic blackbox optimization solvers \cite{custodio2017,chang2023b}.

In such applications, a set of design points is generated and evaluated (through simulation or experimentation) to produce an initial data set. This data set is used to fit a surrogate model of the underlying blackbox process, which is then used for the purpose of approximation or optimization in downstream applications. The accuracy of the resulting approximation or global convergence of the optimization is greatly affected by the quality of the initial design of experiments.

To facilitate the initial design, a significant amount of research \cite{garud2017,joe2008,johnson1990} and software \cite{attia2023,lee2015,virtanen2020,wang2022}
are dedicated to generating space-filling experimental designs and intelligent sampling techniques. While approaches may differ based on specific applications, in the context of surrogate modeling and design optimization for deterministic processes, a space-filling design will constitute data sample locations. These locations are typically obtained from a simply-bounded subset of $\mathbb{R}^d$, which we will refer to as the {\sl design space}. Then the problem is that of generating data samples that  guarantee accuracy and representation across a large percentage of the design space.  One challenge for generating design of experiments is that the deterministic techniques such as those in \cite[Ch.\ 3 \& 4]{myers2016} require exponentially many samples with increasing dimension. 

It is known that the need for computational resources increases exponentially with the number of samples and the dimensionality of data. Although, with the advent of high performance computers, it is sometimes possible to fulfill the computational necessity for moderate dimensions, the concentration of measure still prevalent in high dimensions is rather tricky to handle~\cite{gorban2017}. Therefore,  favorable properties in high-dimensions is a prime requirement in generation of space-filling designs. These designs are often based upon quasi-random and low-discrepancy sequences that attempt to produce random samples .  While these techniques are cheap and effective, they only address the fundamental challenge of measure collapse in high-dimensions heuristically.

In this paper, we propose a novel technique (SF-SFD) for generating high-dimensional distribution functions. Our technique is designed to generate probability distribution functions (pdfs) that are as robust as possible against measure collapse.
One can sample from these pdfs to produce high-dimensional space-filling designs with high probability.
To achieve this, we will optimize the Fourier coefficients of the pdf in order to minimize the expected statistical discrepancy of each sample.
This approach directly addresses the collapse of the underlying distribution and produces better space-filling designs in high dimensions than other randomized methods on a limited budget.

The  remainder of this paper is organized as follows.
In Section~\ref{sec:background} we will provide additional information on techniques for design of experiments and metrics for assessing their quality.
We conclude that for the class of problems that we are interested in, randomized and quasi-randomized methods are most appropriate.
In Section~\ref{sec:measure} we will explain how a concentration of measure makes randomized techniques ineffective for high-dimensional problems.
In Section~\ref{sec:methods} we will introduce a novel method for tuning pdfs in order to slow the concentration of measure.
In Section~\ref{sec:results} we will provide some initial results showing that our method can succeed beyond random sampling, and even becomes more effective than other state-of-the-art techniques in very high-dimensional design spaces.
In Section~\ref{sec:conclusion} we will summarize these results and summarize the next steps for this work.

\section{Background}
\label{sec:background}
Existing experimental design methods can be broadly categorized as adaptive methods, which utilize response values when selecting sample points, and non-adaptive methods, which do not.
It is well-known that {\sl adaptive} search methods are often more efficient in practice, when only considering the accuracy in solving a given task as a function of the number of samples taken.
Well-known adaptive search techniques include, DIRECT \cite{jones1993}, Bayesian optimization \cite{garnett2023}, and various forms of active learning \cite{sapsis2022}.
However, {\sl non-adaptive} methods are useful in situations where large batches are needed (e.g., pre-planning batched chemistry experiments); can be used to multi-start local modeling and optimization techniques; and are used to initialize many adaptive techniques (e.g., Latin hypercubes to start Bayesian optimization).
Therefore, our focus in this paper is limited to non-adaptive methods, which should be considered separately.

The goal of non-adaptive methods in the space-filling design setting is to obtain a design ${\cal X}$ consisting of $n$ points from a simply-bounded $d-$dimensional region of $\mathbb{R}$. Without loss of generality, we assume that we are sampling from the unit cube $[0, 1]^d$.
Since the response values of the samples are not available at this time, the utility of a sample is measured purely by how well it fills the space.
Therefore, it should (approximately) solve
\begin{equation}
    \max_{\{ {\cal X} : {\cal X} \subset [0,1]^d, |{\cal X}| = n\}} T({\cal X}) \quad\text{ or}\quad 
    \min_{\{ {\cal X} : {\cal X} \subset [0,1]^d, |{\cal X}| = n\}} T({\cal X}),
\end{equation}
where $T(\cdot)$ is a design optimality criteria,
and $n$ is the sample size. The nature of $T(\cdot)$  determines whether the problem is to maximize or minimize.
Common examples of design optimality criteria $T(\cdot)$ include
\begin{enumerate}
    \item the discrepancy of the sample (should be minimized) \cite{joe2008},
    \item the A, E, or D-optimality score of the information matrix (should be maximized) \cite{attia2023,wang2022}, and
    \item geometry-based criteria such as maximin (maximized) and minimax (minimized) distance criteria \cite{johnson1990,pronzato2017}.
\end{enumerate}
In this work, the sample size $n$ will typically be in the hundreds. There are numerous applications where $n$ may be larger or smaller, but this is an appropriate range for many real-world design optimization problems such as those listed in Section~\ref{sec:intro} \cite{chang2023b,custodio2017,neveu2022,wang2023}.

The A, E, and D optimality criteria are based on information theory, and generally require access to the Fisher information matrix.
Therefore, many existing techniques for generating samples based on these criteria are model-based and not applicable for our blackbox setting \cite{wang2023}.
Geometric distance-based criteria, such as maximin distance, can be related to properties of the Fisher information matrix, but are easier to compute for a generic surrogate modeling method.
However, such designs can still be combinatorially expensive to generate in high dimensions due to their connection to Delaunay triangulations \cite{pronzato2017}.
All of these optimality criteria are commonly used in the field of optimal experimental design, where the location of each design point is posed as a variable in an optimization problem \cite{attia2023}. Although these techniques are appropriate when $n$ is small in applications such as sensor placement, these techniques are difficult to scale for our target application.

In this paper we will mainly focus on capturing the effect of measure concentration. While additional details about measure concentration can be found in section~\ref{sec:measure}, in brief, we are concerned with the phenomenon where uniformly distributed random samples congregate to a small region of the sample space. As this behavior reflects through the imbalance in the density function,  discrepancy of a sample is an appropriate metric to capture it~\cite{discrep_book}. Precisely, let ${\cal Y}$ be an infinite sequence of points in $[0,1]^d$, and let ${\cal Y}_N$ denote the first $N$ points in ${\cal Y}$. Then
\begin{equation}
    D_N({\cal Y}) = \sup_{B\in J}\left| \frac{|\{{\bf y_i} : {\bf y_i} \in B \text{ and } y_i \in {\cal Y}_N\}|}{N} - \mu(B) \right|,
    \label{eq:discrep}
\end{equation}
where $\mu(\cdot)$ is the Lebesgue measure in $\mathbb{R}^d$ and $J$ is the set of all Lebesgue measurable subsets of $[0, 1]^d$.

The sequence ${\cal Y}$ is said to be low-discrepancy if $\lim_{N\rightarrow\infty}D_N({\cal Y}) = 0$, and the discrepancy $D_N(\cdot)$ is often used as a measure of a finite sample's uniformity.
When used as a measure of uniformity, a discrepancy $D_n({\cal X})$ that is close to $1$ corresponds to an imbalance in the distribution of ${\cal X}$ in $[0, 1]^d$, while a discrepancy close to $0$ corresponds to a general uniformity of ${\cal X}$ in $[0, 1]^d$.
Since it is desirable to fill all areas of the design region when sampling, low-discrepancy samples are considered better, and many optimization libraries use low-discrepancy sequences, such as the Sobol sequence \cite{balandat2020,custodio2017} as a substitute for uniform-random sampling.

Moreover, as it is impossible to compute the exact discrepancy as it is defined in (\ref{eq:discrep}), various approximations of the discrepancy are utilized in the literature. One may refer to \cite{discrep_book} for further details. We use $\mathcal{L}_2$ discrepancy for our approach, which is a special case of the $\mathcal{L}_p$ discrepancy defined as
\begin{equation*}
    \mathcal{L}_p({\cal X}):= \left ( \int_{[0,1]^d} \left | \frac{\left|{\cal X} \cap [0,\mathbf{x}) \right|}{N}  - \mu\left([0,\mathbf{x})\right)\right |^p d \mathbf{x} \right )^{(1/p)}
\end{equation*}
where $1\le p < \infty$.

In this work, we use the centered ${\cal L}_2$ discrepancy of \cite{hickernell1998}
\begin{align*}
{\cal L}_{2}^{C}({\cal X}) =& \left( \frac{13}{12}\right)^d - 
\frac{2}{n} \sum_{i=1}^n \prod_{k=1}^d \left(1 + \frac{1}{2}\left|x_{i,k} - 0.5\right| - \frac{1}{2}\left|x_{i,k} - 0.5\right|^2 \right) \\
& + 
\frac{1}{n^2} \sum_{i=1}^n \sum_{j=1}^n \prod_{k=1}^d \left(1 + \frac{1}{2}\left|x_{i,k} - 0.5\right| + \frac{1}{2}\left|x_{j,k} - 0.5\right| - \frac{1}{2}\left|x_{i,k} - x_{j,k}\right|^2 \right)
\end{align*}
where $x_i$ and $x_j$ are points in ${\cal X}$, and $x_{i,k}$ indicates the $k$th component of $x_i$.
Note that although the discrepancy should take a value between $0$ and $1$, ${\cal L}_2^C$ is an approximation based on numerical quadrature, which can take on much larger values for ill-spaced samples.
In particular, note that when all $x_i\in{\cal X}$ are clustered near the center of the design region $[0, 1]^d$ (as is often the case during measure collapse), the value of ${\cal L}_2^C({\cal X})$ may approach $\left(\frac{13}{12}\right)^d - 1$.
This is an important observation since we will observe large values of ${\cal L}_2^C({\cal X})$ in Section~\ref{sec:results}.

Many techniques have been implemented to produce samples of low discrepancy \cite{sobol1967,wong1997}. %
In practice, the randomized Sobol sequence \cite{paul1988,joe2003,joe2008} is most commonly used in RSM applications \cite{custodio2017}.
However, for the Sobol sequence and other low-discrepancy sequences, performance can still degrade drastically for sample sizes that are overly small with respect to the dimension, or for sample sizes that are not multiples of some preferred size (e.g., powers of 2).

Therefore, many RSM applications use the heuristic of Latin hypercube sampling \cite{neveu2022,chang2023b,muller2017}.
This is even typically recommended as a means to start adaptive sampling techniques, such as Bayesian optimization \cite[Ch.\ 9.3]{garnett2023}.
While Latin hypercube sampling is effective in practice, it is essentially generates samples that are stratified over a single dimension, which is useful for single-variable analysis.
However, there are no guarantees of uniformity over multiple dimensions, and analyses based on Latin hypercubes may miss multivariate interactions.
Therefore, Latin hypercubes are only considered optimal when they have been optimized with respect to another design optimality criteria, such as A, E, or D optimality, which is a combinatorially hard problem \cite{viana2016}.

\subsection{Summary and Key Challenges}
It is well-known that as the dimension of the space grows, the number of samples needed to construct an accurate statistical, numerical, or machine learning model also grows exponentially.
This challenge is known as the {\sl curse of dimensionality}. 
Therefore, for a fixed sample size, the quality of the surrogate model will reduce with increasing dimension, regardless of the sampling technique used due to a reduction in sample density.

In regimes where $n$ is too large for an optimal experimental design and $d$ is too large for a deterministic design of experiments, applications typically resort to Sobol sequences \cite{custodio2017} or Latin hypercube designs \cite{chang2023b,muller2017}.
Both of these techniques are generally seen as approximations to uniform random sampling with better high-dimensional properties.
However, none of these directly address the fundamental issue, which is the collapse of the measure in high dimensions.

In our case, when a sample point $x_i$ is drawn uniformly from $[0, 1]^d$, its squared distance to the center of the design space
is given by $\|x_i - \frac{1}{2}\|_2^2 = \sum_{k=1}^d{(x_{i,k} - \frac{1}{2})^2}$.
So for each component $k=1$, $\ldots$, $d$,
$$
\mathbb{E}\left[ \left(x_{i,k} - \frac{1}{2}\right)^2 \right]
= \int_{0}^1 \left(x - \frac{1}{2}\right)^2 dx 
= \frac{1}{3} - \frac{1}{2} + \frac{1}{4}
= \frac{1}{12}
$$
and the variance of the expected value is a finite constant $\nu$.

Thus, by the central limit theorem (CLT), for all $x_i \in {\cal X}$,
$\mathbb{E}[\|x_i - \frac{1}{2}\|_2^2] = \frac{d}{12}$ with variance $\frac{\nu}{d}$.
So, $x_i$ will concentrate on a sphere of radius $\sqrt{\frac{d}{12}}$ centered at $(0.5, \ldots, 0.5)^\top$ with vanishing standard deviation as $d$ increases.
For large values of $d$, this will leave both the center and corners of the design space empty.

This motivates the need for a technique that directly addresses this issue of measure collapse and scales to larger samples in moderate to high dimensional design spaces. In the following section, we will explain how the measure collapses for high-dimensional samples and what can be done to address this issue in the context of sampling for surrogate modeling.

\section{Concentration of Measure in High Dimensions}
\label{sec:measure}
Consider a random variable $X$ in $\mathbb{R}^{d}$ and let the probability density function~(pdf) be given as $p(x).$ Say, we seek to write the exact expression for the pdf. Traditionally, one would begin by writing the characteristic function $C(t,X)$ for the random variable $X$ which is achieved by applying a Fourier transform on the random variable $X$ and writing a series expansion on the Fourier transform~\cite{adeniran2020derivation}. Particularly, in the case when $X$ corresponds to continuous Lebesgue measure, the characteristic function is written as 
\begin{align}
    C(t, X) = \mathbb{E} \left[ e^{i t X}\right]  =  \int \left[ e^{i t X}\right] d \mu,
\end{align}
where $\mu \in \mathbb{R}^{d}$ is a Lebesgue measure on a $d$ dimensional vector space. The precise derivation of the density from this point requires evaluating the inversion of the characteristic function through the Levy's inversion formulae~\cite{hewitt1953remarks}. This precise expression is given as  
\begin{align}\label{eq:inversion}
    p (x) = \int e^{-i t \mathbb{E} \left[ e^{i t X} \right] } dt,
\end{align}
where $p (x)$ describes the probability density function~(pdf). Solving the integral on the right hand side would provide the pdf of Gaussian.   To solve the integral \eqref{eq:inversion},  we must expand $e^{-i t \mathbb{E} \left[ e^{i t X} \right] }$ expansion applied on the expected value of  $e^{i t X}.$ Therefore,  the quality of the pdf estimation depends on the availability of samples from $X.$  In the scenario when $d \rightarrow \infty,$ the volume will concentrate~\cite{gorban2017}, which means that the samples from $X$ will describe a small region from the high dimensional space. In other words, the volume of a cube that was well spread across different areas becomes concentrated towards as a sphere around the origin point as detailed in the previous section.  

This phenomenon leads to two scenarios, first, the estimation of $\mathbb{E} \left[ e^{i t X} \right] $ does not change for different samples of $X$ as the concentrated volume will provide the same samples over and over again. Therefore, the inversion will provide the same density values for different samples assuming that the inverse is well defined. Second, the concentration phenomenon leads to a situation where the presence of many zeros or close to zero values in the data samples will introduces zero modes in $\mathbb{E} \left[ e^{i t X} \right].$ In the matrix sense, many of the eigenvalues of $\mathbb{E} \left[ e^{i t X} \right]$ will end up being zero and will lead to singular modes in the density. That is, the {\sl empirical} density function will be ill-defined in many regions of the design space.   These two issues prevalent in high dimensions will prohibit the use to uniform sampling  because uniform sampling from the original design space will end up with an ill-defined empirical distribution function.

An alternative avenue is to generate samples more intelligently by considering the density rather than blind uniform sampling. Towards this end, we will not solve the problem by sampling-based likelihood estimation like a Monte Carlo approach, but, rather construct a distribution function that is optimized through an iterative procedure. This distribution function is an approximation of inverse map $M^{-1} \mathbb{E} \left[ e^{i t X}, \right],$ on the design space. In particular, we will write the pdf as a linear combination of individual terms of the Fourier series with coefficients. Then, we will solve for the coefficients through an iterative optimization approach. By successively deriving samples and the corresponding pdf, we will attempt to find the approximated pdf that best explains the distribution of the data. It is our hypothesis that by optimizing the distribution function instead of the likelihood we will sidestep the impact of measure concentration on sample and the density will be better defined. In what follows next, we will detail our approach and later describe the advantages of our method in this domain through a simulation study.

\section{Our approach}
\label{sec:methods}

Now that we have established the issue with existing sampling techniques and the issue of measure collapse, we are ready to propose our solution, which we refer to as the Stochastic Fourier space-filling design (SF-SFD).  In this approach, we create a sample, ${\cal X}$, of any size $n$ from a simply bounded $d-$dimensional space $[0,1]^d \subset \mathbb{R}^d$. We achieve this by optimizing the Fourier coefficients of the pdf, in order to compensate for measure collapse at the specified values of $n$ and $d$.

To measure the collapse of the pdf, we use the centered ${\cal L}_2$ discrepancy, ${\cal L}_2^C({\cal X})$ from Section~\ref{sec:background}.
We argue that the discrepancy is an appropriate choice of performance metrics since large discrepancy values are correlated to measure collapse.
Additionally, the centered ${\cal L}_2$ approximation will fail and produce unrealistically large numbers when points are overly clustered near the origin. Since the radius of our sphere of concentration (see Section \ref{sec:background}) is a small percentage of the total design space, excessively large discrepancies will be indicative of measure collapse.

At first glance, the fact that we are optimizing our design toward well-known optimality criteria could be seen as similar to optimal experimental design \cite{attia2023,wang2023}
or Latin hypercube optimization \cite{viana2016}.
However, we focus on optimizing properties of the underlying pdf to prevent a collapse of measure, rather than focusing on the placement of individual data points.
This is a key difference, and we believe that the proposed technique will be more scalable while also directly addressing the underlying issue suffered by na{\"ive} randomized techniques.
This strategy is motivated by the connection between measure collapse and singularity of the Fourier transform of the pdf, as discussed in Section~\ref{sec:measure}.

Our approach can be summarized by the following $3$-step process:
\begin{enumerate}
\item We create an initial probability distribution function for SF-SFD based on a uniform distribution.
\item We take a discrete Fourier transform (DFT) of the square-roots of the probabilities to obtain tunable coefficients.
See Section~\ref{sec:coeffs} for further details.
\item Since the DFT is a unitary operator, we can take perturbations on the surface of the unit sphere in order to generate new (square-root) probability density functions.
We will use a constrained optimization procedure to iteratively generate perturbations to our Fourier coefficients with the objective of minimizing the expected {\sl empirically observed} discrepancy of the resulting pdf.
See Section~\ref{sec:optimization} for more details.
\end{enumerate}
The process described above is outlined in Algorithm 1, with further details in Sections~\ref{sec:coeffs} and \ref{sec:optimization}.

\begin{algorithm}
    \caption{SF-SFD}
    \label{alg:sf-sfd}
    \begin{algorithmic}
    \State Let $P^{(1,m)}$ denote the current 1D pdf and $Q^{(1,m)}$ denote $\sqrt{P^{(1,m)}}$, as described in Section~\ref{sec:coeffs};
    \State Let $C^{(1,m)}$ denote the complex-valued FFT of $Q^{(1,m)}$ as described in Section~\ref{sec:coeffs};
    \State Let $\theta$ denote the optimization variables, as described in Section~\ref{sec:optimization};
    \State $a_i$ is the current number of draws to estimate expected-value, as described in Section~\ref{sec:optimization};
    \State \textbf{Initialize} $P^{(1,m)} = $ 1D uniform distribution; $Q^{(1,m)}$, $C^{(1,m)}$, and $\theta$ are set accordingly;
    \While{optimization stop conditions not met}
    \State Optimizer iterates to generate perturbation $C'^{(1,m)}$ to $C^{(1, m)}$;
    \State Reverse the process from Section~\ref{sec:coeffs} to recover the perturbed 1D pdf $P'^{(1,m)}$;
    \State Estimate expected discrepancy by drawing $a_i$ iid $d$-dimensional samples of size $n$ from $P'^{(1,m)}$;
    \State For the next iteration, update: $P^{(1,m)} \leftarrow P'^{(1,m)}$;
    \State Update $Q^{(1,m)}$, $C^{(1,m)}$, and $\theta$ accordingly;
    \State Increment $a_i$ if needed;
    \State Return the estimated expected discrepancy to the optimizer for the next iteration;
    \EndWhile
    \end{algorithmic}
\end{algorithm}

\subsection{Obtaining the Fourier coefficients}
\label{sec:coeffs}
In Step 1 of our three-step process, we start from a uniform distribution on $[0, 1]^d$.
Ideally, we would like to optimize the Fourier transform of this $d$-dimensional distribution function $P^{(d)}$, but this would be computationally intractable for large values of $d$. Instead, we make the simplifying assumption that our distribution function will always be symmetrical in each dimension (in other words, the projection onto each coordinate axis is identical). This allows us to instead train a one-dimensional distribution function $P^{(1)}$ and draw $d$ i.i.d.\ samples from it to obtain a single $d$-dimensional design point. In order to make the optimization problem finite, we discretize $P^{(1)}$ into $m$ discrete cells of equal mass $p_1$, $\ldots$, $p_m$. This defines our one-dimensional probability mass function $P^{(1,m)}$.

Next, let $Q^{(1,m)} = \{q_i\}_{i=1}^m$, where $q_i = \sqrt{p_i}$ for $i=1$, $\ldots$, $m$. This is the square-root probability mass function, upon which we now apply a DFT. In this work, we use the unitary form of the DFT such that
the 2-norms are preserved in the Fourier space and through the inverse Fourier transform (IFT) by Parseval's theorem \cite[Theorem 21]{Lax2002}.
Since $ \sum_{i=0}^{m} |q_i|^2 = 1$, it implies $\sum_{i=0}^{m} |c_i|^2 = 1$ where $C^{(1,m)} =
\begin{bmatrix}
c_0 & c_1  &\cdots  & c_{m}
\end{bmatrix}$
are the complex-valued Fourier coefficients. We can then perturb the coefficients $c_i ~\text{for}~ i \in \{0, \ldots, m\}$ to obtain $c_i'$ such that $\sum_{i=0}^{m} |c_i^{'}|^2 = 1$ is still satisfied.
Once we have the new coefficients for $C'^{(1,m)}$, we can invert the above process to obtain the updated square-root mass function $Q'^{(1,m)}$, and eventually our updated one-dimensional mass function $P'^{(1,m)}$.
Since we are assuming that the distribution function is symmetric in all dimensions, we can draw a $d$-dimensional sample of size $n$ via $d \times n$ i.i.d.\ draws from $P'^{(1,m)}$ to sample the perturbed $d$-dimensional pdf.

\subsection{Optimization of Fourier Coefficients}
\label{sec:optimization}
In order to tune our pdf using optimization (Step 3 above), we can introduce any perturbation to the Fourier coefficients, so long as the coefficients have a unit 2-norm. These perturbations on $c_i$ can be chosen to specifically decrease $\mathbb{E}\left[{\cal L}_2^C({\cal X})\right]$.
In order to tune this criteria while maintaining the requirement that $\sum_{i=0}^{m} |c_i^{'}|^2 = 1$, we must solve an optimization problem on a $(m-1)$-dimensional complex unit sphere (Bloch sphere).

To put this concisely, we are solving the complex-valued optimization problem
$$
\min_{C^{(1,m)} \in \mathbb{C}^m} \mathbb{E}\left[ L_2^C(X) | X \sim P^{(d,m)}; \sum_i |c_i|^2 = 1 \right]
$$
where $P^{(d,m)}$ is a $d$-dimensional pdf with implicit dependence upon $C^{(1,m)}$, via the process from Section~\ref{sec:coeffs}.

In practice, it is not easy to solve a complex-valued optimization problem with nonlinear (spherical) constraints.
Thus, we represent $c_i$ in polar form eliminating the need for the spherical constraint and complex variables.
Specifically, we generate the real-valued Euler angles $\theta_i$, $i=1$, $\ldots$, $2m-1$, which point to coordinates on the $m$-dimensional complex sphere.
These angles can be optimized by a blackbox solver subject only to the linear constraints $0 \leq \theta_i \leq 2\pi$, $i=1$, $\ldots$, $2m-1$.
Again, note that because we assume symmetry in all dimensions, the dimension of the optimization problem depends linearly on $m$, but is independent of both $d$ and $n$.

Recall that we are attempting to generate a distribution function that can be used to produce good samples of size $n$ in $[0, 1]^d$. Thus, our objectives are based on the expected performance of a realized sample of a pre-specified size $n$ from the resulting distribution. Note that the above is a stochastic problem since there will be variation in each individual design drawn from our pdf.
In order to address this problem, in this work, we estimate the expected value of each performance measure based on an average over $a_i$ designs of size $n$ drawn from the resulting distribution.
In order to guarantee convergence to the true expected value, we gradually increase our sample size $a_i$ with the iteration index $i$. In this work, we start with a sample size of $a_1 = 50$ and increase $a_i$ by $1$ every $10$ iterations. 
This guarantees that the error term in the stochastic approximation (SA) vanishes in the limit, which is a necessary condition to the convergence of SA methods \cite{lai2003}.

Now that we have posed a bound-constrained optimization problem of moderate dimension, we 
calculate ${\cal L}_2^C{\cal X}$ using the implementation in {\tt scipy.stats.qmc} \cite{hickernell1998,virtanen2020}.
Then we use the COBYLA implementation in {\tt scipy.optimize.minimize} for solving this bound-constrained blackbox optimization problem \cite{powell1994,virtanen2020}.
Although COBYLA was originally proposed as a deterministic solver, due to its similarities to stochastic gradient descent, COBYLA is known to perform very well for stochastic problems such as this \cite{shi2021}.

\section{Results}
\label{sec:results}

To test our method, we compare the expected discrepancy of our final pdf against the expected discrepancy of a Latin hypercube sample (LHS), Sobol sequence (Sobol), and random sample of size $n$.
In the case of SF-SFD, we have optimized our pdf using the exact methods described in Section~\ref{sec:methods}.
For the discretization of our mass function, we used a value of $m=10$, and for COBYLA we use the default setting in {\tt scipy} \cite{virtanen2020}.
For the comparisons, we have used the Latin hypercube sampling and Sobol sequence implementations from {\tt scipy.stats.qmc} \cite{Roy2023}.
In every case, we average results over 10 distinct random seeds.

\begin{figure}[h]
    \centering
    \includegraphics[width=0.49\textwidth]{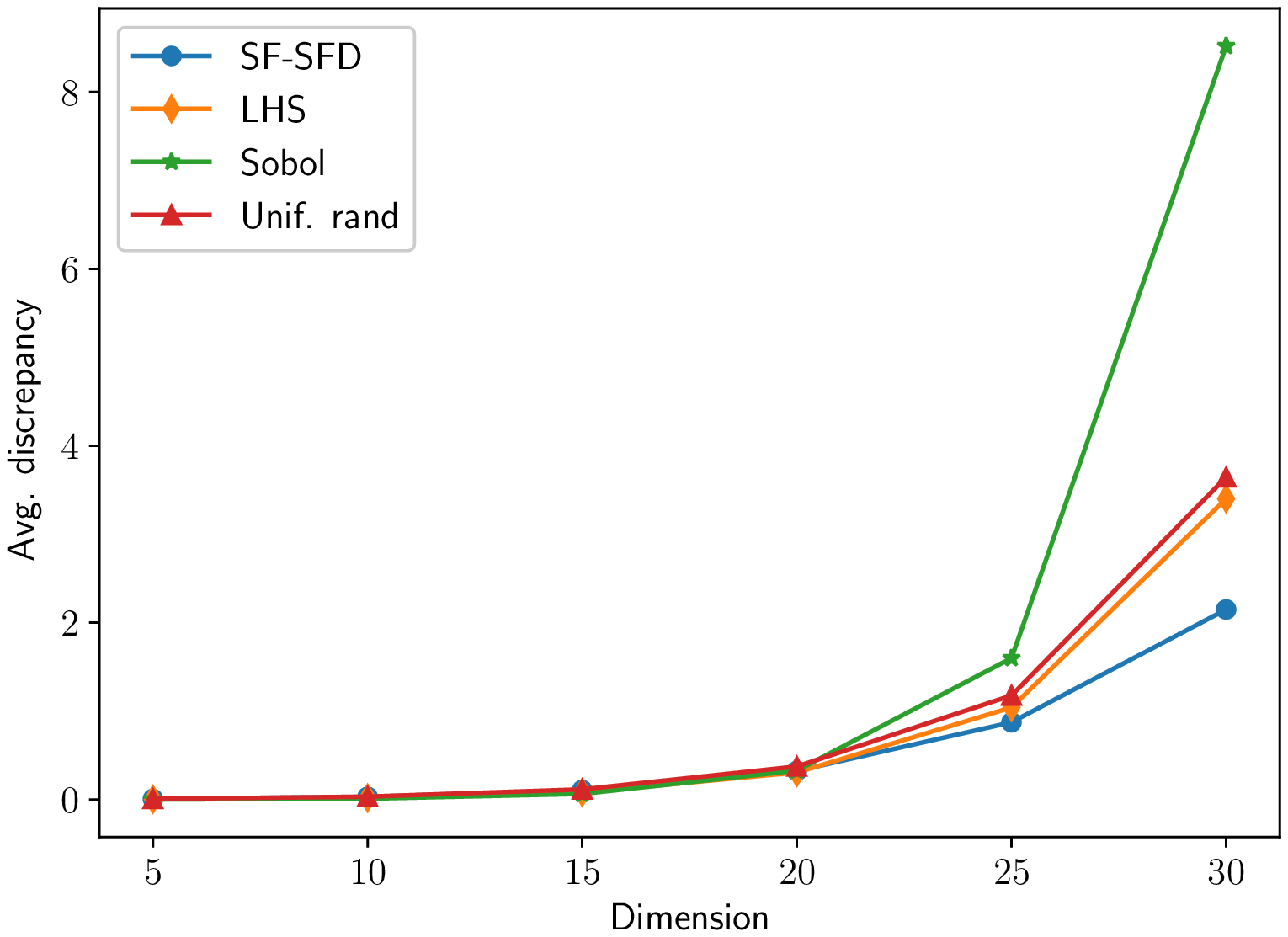}    \includegraphics[width=0.49\textwidth]{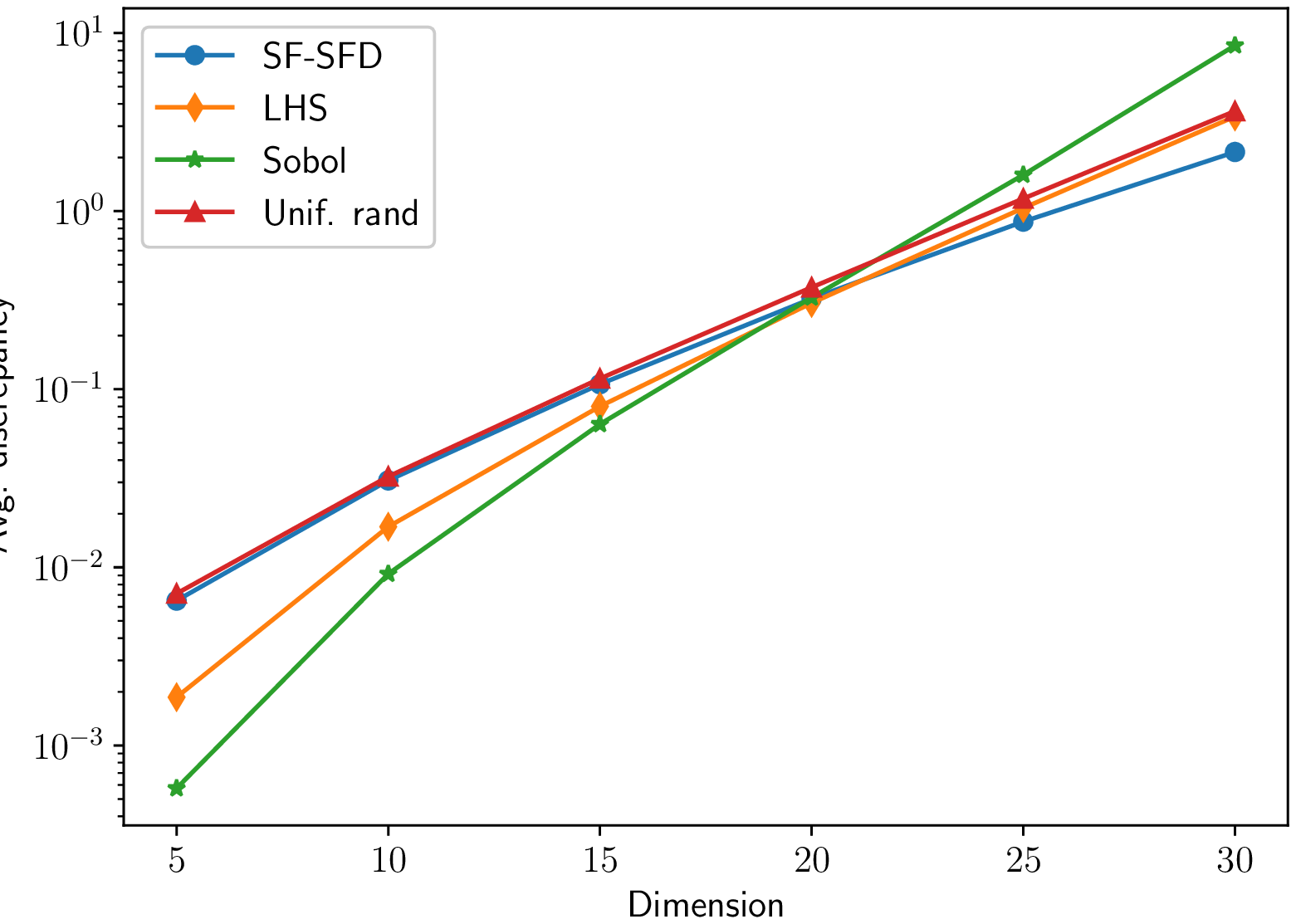}
    \caption{Average discrepancy of generated samples with increasing dimension at linear (left) and logarithmic (right) scales.}
    \label{fig:discrep-results}
\end{figure}

We have performed experiments with sample sizes of $n=100$, $200$, $300$, $400$, and $500$, and $d=5$, $10$, $15$, $20$, $25$, and $30$.
The averaged discrepancies for all four methods at all problem sizes are shown in Table~\ref{tab:discrep} and the increase in discrepancy with $d$ (averaging over all values of $n$) is plotted in Figure~\ref{fig:discrep-results}.

\begin{table}[h]
    \centering
\begin{tabular}{|c|l|ccccc|l}
\cline{1-7}
\multirow{2}{*}{\textbf{Dimension}} & \multicolumn{1}{c|}{\multirow{2}{*}{\textbf{$~$ Method $~$}}}
                                      & \multicolumn{5}{c|}{\textbf{Sample sizes}}  &  \\
\cline{3-7} & \multicolumn{1}{c|}{} & \multicolumn{1}{c|}{$~~~~~\mathbf{100} ~~~~~$}
                                      & \multicolumn{1}{c|}{$~~~~~\mathbf{200} ~~~~~$}
                                      & \multicolumn{1}{c|}{$~~~~~~\mathbf{300} ~~~~~$}
                                      & \multicolumn{1}{c|}{$~~~~~\mathbf{400} ~~~~~$} &
                                      $~~~~~\mathbf{500} ~~~~~$ &  \\
\cline{1-7}

\multirow{4}{*}{5}
    & SF-SFD & \multicolumn{1}{c|}{0.0142}     & \multicolumn{1}{c|}{0.0071}
             & \multicolumn{1}{c|}{0.0047}     & \multicolumn{1}{c|}{0.0036}
             & 0.0029     &  \\ \cline{2-7}
    & LHS    & \multicolumn{1}{c|}{0.0042}     & \multicolumn{1}{c|}{0.0020}
             & \multicolumn{1}{c|}{0.0014}     & \multicolumn{1}{c|}{0.0010}
             & 0.0008     &  \\ \cline{2-7}
    & Sobol  & \multicolumn{1}{c|}{\bf 0.0017}     & \multicolumn{1}{c|}{\bf 0.0006}
             & \multicolumn{1}{c|}{\bf 0.0003}     & \multicolumn{1}{c|}{\bf 0.0002}
             & \bf 0.0001     &  \\ \cline{2-7}
    & Unif.\ Rand.\ & \multicolumn{1}{c|}{0.0157}     & \multicolumn{1}{c|}{0.0078}
                    & \multicolumn{1}{c|}{0.0052}     & \multicolumn{1}{c|}{0.0039}
                    & 0.0031     &  \\ \cline{1-7}

\multirow{4}{*}{10}
    & SF-SFD & \multicolumn{1}{c|}{0.0670}     & \multicolumn{1}{c|}{0.0335}
             & \multicolumn{1}{c|}{0.0226}     & \multicolumn{1}{c|}{0.0170}
             & 0.0136     &  \\ \cline{2-7}
    & LHS    & \multicolumn{1}{c|}{0.0376}     & \multicolumn{1}{c|}{0.0179}
             & \multicolumn{1}{c|}{0.0125}     & \multicolumn{1}{c|}{0.0091}
             & 0.0073     &  \\ \cline{2-7}
    & Sobol  & \multicolumn{1}{c|}{\bf 0.0240}     & \multicolumn{1}{c|}{\bf 0.0100}
             & \multicolumn{1}{c|}{\bf 0.0058}     & \multicolumn{1}{c|}{\bf 0.0037}
             & \bf 0.0025     &  \\ \cline{2-7}
    & Unif.\ Rand.\ & \multicolumn{1}{c|}{0.0710}     & \multicolumn{1}{c|}{0.0353}
                    & \multicolumn{1}{c|}{0.0236}     & \multicolumn{1}{c|}{0.0177}
                    & 0.0141     &  \\ \cline{1-7}

\multirow{4}{*}{15}
    & SF-SFD & \multicolumn{1}{c|}{0.2259}     & \multicolumn{1}{c|}{0.1183}
             & \multicolumn{1}{c|}{0.0801}     & \multicolumn{1}{c|}{0.0607}
             & 0.0487     &  \\ \cline{2-7}
    & LHS    & \multicolumn{1}{c|}{0.1765}     & \multicolumn{1}{c|}{0.0870}
             & \multicolumn{1}{c|}{0.0578}     & \multicolumn{1}{c|}{0.0444}
             & 0.0348     &  \\ \cline{2-7}
    & Sobol  & \multicolumn{1}{c|}{\bf 0.1645}     & \multicolumn{1}{c|}{\bf 0.0659}
             & \multicolumn{1}{c|}{\bf 0.0395}     & \multicolumn{1}{c|}{\bf 0.0276}
             & \bf 0.0207     &  \\ \cline{2-7}
    & Unif.\ Rand.\ & \multicolumn{1}{c|}{0.2520}     & \multicolumn{1}{c|}{0.1255}
                    & \multicolumn{1}{c|}{0.0834}     & \multicolumn{1}{c|}{0.0627}
                    & 0.0501     &  \\ \cline{1-7}

\multirow{4}{*}{20}
    & SF-SFD & \multicolumn{1}{c|}{\bf 0.6557}     & \multicolumn{1}{c|}{0.3628}
             & \multicolumn{1}{c|}{0.2500}     & \multicolumn{1}{c|}{0.1907}
             & 0.1549     &  \\ \cline{2-7}
    & LHS    & \multicolumn{1}{c|}{0.6688}     & \multicolumn{1}{c|}{0.3357}
             & \multicolumn{1}{c|}{0.2212}     & \multicolumn{1}{c|}{0.1665}
             & 0.1330     &  \\ \cline{2-7}
    & Sobol  & \multicolumn{1}{c|}{0.8618}     & \multicolumn{1}{c|}{\bf 0.3331}
             & \multicolumn{1}{c|}{\bf 0.1979}     & \multicolumn{1}{c|}{\bf 0.1386}
             & \bf 0.1052     &  \\ \cline{2-7}
    & Unif.\ Rand.\ & \multicolumn{1}{c|}{0.8171}     & \multicolumn{1}{c|}{0.4089}
                    & \multicolumn{1}{c|}{0.2723}     & \multicolumn{1}{c|}{0.2044}
                    & 0.1636     &  \\ \cline{1-7}

\multirow{4}{*}{25}
    & SF-SFD & \multicolumn{1}{c|}{\bf 1.6263}     & \multicolumn{1}{c|}{\bf 0.9964}
             & \multicolumn{1}{c|}{\bf 0.7221}     & \multicolumn{1}{c|}{\bf 0.5621}
             & 0.4591     &  \\ \cline{2-7}
    & LHS    & \multicolumn{1}{c|}{2.2569}     & \multicolumn{1}{c|}{1.1541}
             & \multicolumn{1}{c|}{0.7625}     & \multicolumn{1}{c|}{0.5745}
             & 0.4564     &  \\ \cline{2-7}
    & Sobol  & \multicolumn{1}{c|}{4.4776}     & \multicolumn{1}{c|}{1.5757}
             & \multicolumn{1}{c|}{0.8858}     & \multicolumn{1}{c|}{0.6033}
             & \bf 0.4514     &  \\ \cline{2-7}
    & Unif.\ Rand.\ & \multicolumn{1}{c|}{2.5771}     & \multicolumn{1}{c|}{1.2875}
                    & \multicolumn{1}{c|}{0.8578}     & \multicolumn{1}{c|}{0.6439}
                    & 0.5154     &  \\ \cline{1-7}

\multirow{4}{*}{30}
    & SF-SFD & \multicolumn{1}{c|}{\bf 3.7137}     & \multicolumn{1}{c|}{\bf 2.4248}
             & \multicolumn{1}{c|}{\bf 1.8572}     & \multicolumn{1}{c|}{\bf 1.4976}
             & \bf 1.2504     &  \\ \cline{2-7}
    & LHS    & \multicolumn{1}{c|}{7.3923}     & \multicolumn{1}{c|}{3.7545}
             & \multicolumn{1}{c|}{2.4905}     & \multicolumn{1}{c|}{1.8693}
             & 1.4979     &  \\ \cline{2-7}
    & Sobol  & \multicolumn{1}{c|}{25.5500}     & \multicolumn{1}{c|}{8.0504}
             & \multicolumn{1}{c|}{4.2211}     & \multicolumn{1}{c|}{2.7673}
             & 2.0118     &  \\ \cline{2-7}
    & Unif.\ Rand.\ & \multicolumn{1}{c|}{7.9939}     & \multicolumn{1}{c|}{3.9847}
                    & \multicolumn{1}{c|}{2.6560}     & \multicolumn{1}{c|}{1.9928}
                    & 1.5952     &  \\ \cline{1-7}

\end{tabular}
    \caption{Empirical expected value of ${\cal L}_2^C({\cal X})$ averaged over 10 random seeds for SF-SFD, Latin hypercube sampling (LHS), Sobol sequences, and uniform random sampling at various dimensions ($d$) and sample sizes ($n$).
    All values are rounded to 4 decimal places.
    The best performing method at each problem size is emphasized in bold.
    }
    \label{tab:discrep}
\end{table}

As seen in Figure~\ref{fig:discrep-results}, in low dimensions our method performs similarly to random sampling for the tested values of $n$.
On the other hand, the Sobol sequence and Latin hypercube samples (LHS) are significantly better.
As the dimension increases, LHS becomes increasingly similar to uniform random sampling, while the Sobol sequence eventually becomes worse than random sampling as the value of $n$ becomes too sparse with respect to $d$.
On the other hand, our method performs significantly better than uniform random sampling in high dimensions, and greatly slows the rate of measure collapse.
Based on these results, for values of $n$ in the low hundreds, our method is preferable when $d\geq \sim 20$.
In Table~\ref{tab:discrep} we see that SF-SFD overtakes LHS and Sobol sooner for small values of $n$.

It is worth noting that at these problem sizes, the ${\cal L}_2^C$ discrepancy approximation becomes extremely inaccurate due to the bunching of samples near the origin, leaving the corners of the $[0, 1]^d$ completely empty. This is reflected in discrepancy approximations that well-exceed one.
Even our method suffers from this issue, but we are able to mitigate the issue in comparison to other techniques. In general, although the measure concentration may be impossible to avoid, we are able to slow the rate of concentration significantly through SF-SFD's optimization procedure.

\section{Conclusion and Future Work}
\label{sec:conclusion}

In this paper we have proposed a novel method, which we call SF-SFD, for tuning distribution functions in high-dimensional spaces in order to prevent concentration of measure for a finite sample.
We argue that this technique directly addresses the issue of measure concentration and scales better to large dimensions than existing heuristic techniques such as Latin hypercube samples and low-discrepancy sequences such as the randomized Sobol sequence.
We use SF-SFD to generate space-filling design at several common problem dimensions, and show that the average ${\cal L}_2^C$ discrepancy for our designs grows more slowly than other techniques as the dimension of the problem becomes extremely large. The experimental results and analysis presented in this paper can be reproduced by accessing the corresponding GitHub repository at \url{https://github.com/sfdsampling/sfsfd}. The repository contains the necessary code, datasets, and instructions to replicate the experiments and generate the reported results.

Although initially it is difficult to compete with quasi-random methods such as Sobol sequences or even Latin hypercubes, for dimensions exceeding twenty, sample sizes of $100-500$ are not large enough to guarantee reasonable discrepancies with these methods.
In fact, for the ${\cal L}_2^C$ discrepancy approximation used in this paper, samples become so concentrated that the discrepancy estimates begin to blow up beyond the reasonable range.
While all methods are affected, ours is able to slow this rate of concentration as much as possible, achieving better performance for sparse data samples in high dimensions.
It is worth noting that although we have only shown samples of size $100-500$ in dimensions $20-30$ to be sparse enough to warrant our method, every finite sample size will become relatively sparse in sufficiently high dimensions due to the curse of dimensionality.

The next step for this work is to also consider other design optimality criteria, such as A, E, and D optimality and maximin or minimax distances in the formulation of our objective.
While these criteria are not directly related to measure collapse, they are common design optimality criteria in the literature with a direct connection to surrogate model accuracy.
We would also like to prove our method more rigorously by showing that we can obtain a slower rate of measure concentration for our method.
Finally, we will need to show empirically that ``good'' designs generated through our method translate to better approximation performance downstream in real-world applications.

\section*{Acknowledgement}
We would like to thank Stefan Wild and Kesav Krishnan for their feedback regarding an earlier version of this work.

This work was supported in part by the U.S.~Department of Energy, Office of Science, Office of Advanced Scientific Computing Research's SciDAC program under Contract No.\ DE-AC02-06CH11357, the RAPIDS institute 
and by the National Science Foundation
(NSF) Mathematical Sciences Graduate Internship (MSGI) Program sponsored by the NSF Division of Mathematical Sciences. This program is administered by the Oak Ridge Institute for Science
and Education (ORISE) through an inter agency agreement between the U.S. Department of Energy (DOE) and NSF. ORISE is managed for DOE by ORAU. All opinions expressed in this paper are the authors' and do not necessarily reflect the policies and views of NSF, ORAU/ORISE, or DOE.

\small

\bibliographystyle{abbrv}
\bibliography{moo-refs} %

\end{document}